\documentclass[3p,authoryear,times,10pt]{elsarticle}
\usepackage{color}
\usepackage{url}
\usepackage{multirow,setspace,times,amssymb,amsmath,graphicx,color,subfigure,url}
\usepackage{dsfont}
\usepackage{threeparttable}
\usepackage{rotating}

\usepackage{dcolumn}
\usepackage{listings}

\begin{document}

\begin{frontmatter}

\title{Time-dependent lead-lag relationships between the VIX and VIX futures markets}
\author[ECNU]{Yan-Hong Yang \corref{cor1}}
\ead{yhyang@dedu.ecnu.edu.cn}
\author[SSI]{Ying-Hui Shao \corref{cor2}}
\ead{yinghuishao@126.com}

\cortext[cor1]{Correspondence to: 3663 North Zhongshan Road, School of Education, East China Normal University, Shanghai 200062, China.}
\cortext[cor2]{Corresponding author.}
\address[ECNU]{School of Education, East China Normal University, Shanghai 200062, China}
\address[SSI]{School of Statistics and Information, Shanghai University of International Business and Economics, Shanghai 201620, China}

\begin{abstract}
  We utilize the symmetric thermal optimal path (TOPS) method to examine the dynamic interaction patterns between the VIX and VIX futures markets. We document that the VIX dominates the VIX futures more in the first few years, especially before the introduction of VIX options. We further observe that the TOPS paths show an alternate lead-lag relationship instead of a dominance between the VIX and VIX futures in most of the time periods. Meanwhile, we find that the VIX futures have been increasingly more important in the price discovery since the launch of several VIX ETPs.
\end{abstract}
\begin{keyword}
VIX \sep VIX futures \sep Lead-lag relationship \sep Symmetric thermal optimal path
\\
 {\textit{JEL classification:}} C14, G13, G14
\end{keyword}
\end{frontmatter}

\section{Introduction}

The VIX has been the benchmark for stock market volatility, which was introduced by the Chicago Board Options Exchange (CBOE) in 1993 \citep{Frijns2016}. A significant feature of VIX is that VIX tends to be higher when the stock market drops, thus the VIX has also been termed ``investor fear gauge'' \citep{Frijns2016,Zhang2012}. For the purpose of accelerating trade in volatility and hedging, the CBOE launched VIX futures on March 26, 2004 \citep{Frijns2016}. However, trading volume of VIX futures did not show a vigorous growth until the introduction of VIX options on February 24, 2006 \citep{Fernandez-perez2019}. In January 2009, Barclays introduced the first volatility exchange traded product (ETP), termed the VXX. The launch of VIX ETPs really expedite trading in VIX futures \citep{Frijns2016,Fernandez-perez2019}. Furthermore, there has been an booming body of research by both academics and practitioners, into many aspects of VIX and its derivatives \citep{Frijns2016}.

Theoretically speaking, the VIX should neither lead nor lag the VIX futures in accordance with the Efficient Markets Hypothesis \citep{Fama-1970-JF,Fama-1991-JF,Chen2017,Shao2019,Yang2019}. Nonetheless, the complexity and frictions of real markets usually result a mixed lead-lag structure \citep{Shao2019}.
There are several significative study focusing on the lead-lag relationships between the VIX and VIX futures markets. \citet{Zhang2012} explicitly investigate lead-lag dynamics between the VIX and VIX futures, covering the period from March 26, 2004 to May 20, 2009. Via employing a linear Engle-Granger cointegration test with an error correction mechanism, \citet{Zhang2012} uncover that the VIX futures markets have some price-discovery ability. However, adopting a modified Baek and Brock nonlinear Granger test, \cite{Zhang2012} reveal a bi-directional lead-lag interaction between the VIX and VIX futures, suggesting that both VIX and VIX futures react simultaneously to new information. They summarize that the prediction power of VIX futures is unstable and the VIX futures market is information efficient. Correspondingly, \cite{Frijns2016} examine the intraday dynamic interactions of the VIX and VIX futures from January 2, 2008 to December 31, 2014. Using daily data, they provide evidence of price discovery from the VIX futures to the VIX. Moreover, based on ultra-high frequency data, they unveil a bi-directional Granger causality between the VIX and VIX futures. They further document that the VIX futures dominate the VIX more on days with negative returns. Overall, \cite{Frijns2016} conclude that the VIX futures have been increasingly more important to the price discovery process. Likewise, \cite{Bollen2017} suggest that the VIX futures lagged the VIX in the first few years after its introduction, and show a increasing dominance of VIX futures over time. However, \cite{Bollen2017} remark it remains an subtle and open question for the lead-lag relationships between the VIX and VIX futures. Additionally, \cite{Chen2017} reveal that the VIX futures play a dominant role in intraday price discovery.

Indeed, there exist a variety of methods and techniques for detecting the lead-lag structure between spot and futures markets. The broadly adopted approach is the combination of linear and nonlinear Granger causality test \citep{Ashley-Granger-Schmalensee-1980-Em,Bekiros2008}, vector error correction model \citep{Zhang2012}, cointegration test and Markov-switching \citep{Balcilar2015}. Contrast to other popular methods, the TOPS method proposed by \citet{Meng-Xu-Zhou-Sornette-2017-QF} does not need the stationarity of time series. Specifically, this TOPS method is able to clearly characterize the lead-lag relationships among time series, which is demonstrated by intensive synthetic tests in Meng's study \citep{Meng-Xu-Zhou-Sornette-2017-QF}. Subsequently, the TOPS approach has been widely employed to many financial markets, such as the investigation of the interactions between house price and monetary policy \citep{Meng-Xu-Zhou-Sornette-2017-QF}, the research of linkages between crude oil spot and futures markets and so forth \citep{Shao2019}. Given the mixed empirical lead-lag relationships in the research among VIX derivatives, we apply the nonparametric and nonlinear TOPS method to capture the lead-lag structure between the VIX and VIX futures.

The remainder of this paper is organized as follows. Section \ref{Sec: Method} depicts the TOPS method. Section \ref{Sec: Data description} describes the data set and preliminary analysis. Section \ref{Sec: Application} presents applications of TOPS method to the characterization of the dynamic lead-lag relationships between the VIX and VIX futures. Section \ref{Sec: Conclustion} summarizes.

\section{Methodology}
\label{Sec: Method}

\subsection{The symmetric thermal optimal path method}

The TOP method is effective in identifying the structural changes of lead-lag relationships between two time series \citep{Sornette-Zhou-2005-QF,Zhou-Sornette-2006-JMe,Zhou-Sornette-2007-PA}. In 2017, \citet{Meng-Xu-Zhou-Sornette-2017-QF} proposed the TOPS method by introducing the reversal search to the TOP method. Indeed, the TOPS method presents a better performance comparing with the TOP method. The procedure of the TOPS method is depicted as follows.

Considering two time series $X(t_1):t_1=0,\cdots,N-1$ and $Y(t_2):t_2=0,\cdots,N-1$, in which $X(t_1)$ and $Y(t_2)$ represent the standardized logarithmic returns of the VIX futures and VIX, respectively. Thenceforth, we use an $N\times N$ distance matrix $E_{X,Y}$ to comprehensively capture the disparity between $X(t_1)$ and $Y(t_2)$ along a fixed time axis \citep{Shao2019}. The definition of elements of the distance matrix $E_{X,Y}$ is
\begin{equation}
  \epsilon(t_1,t_2) = |X(t_1)-Y(t_2)|~.
  \label{Eq:DistMatrix}
\end{equation}
The mapping $t_1 \rightarrow t_2 = \phi(t_1)$ symbolizes an optimal match between $X(t_1)$ and $Y(t_2)$
\begin{equation}
  \phi(t_1)=\min_{t_2}\{\epsilon(t_1,t_2)\},
  \label{Eq:LocalMinimization}
\end{equation}
where Eq.~(\ref{Eq:LocalMinimization}) represents a local minimization. Considering the unreasonable large jumps, \citet{Sornette-Zhou-2005-QF} derive the lead-lag structure between $X(t_1)$ and $Y(t_2)$ from a global minimization
\begin{equation}
\min \limits_{\{\phi(t_1), t_1=0,1,\ldots,N-1\}} E:= \sum_{t_1=0}^{N-1}|X(t_1)-Y(\phi(t_1))|,
\label{Eq:GlobalMinimization}
\end{equation}
with a continuity constraint
\begin{equation}
0 \leq \phi(t_1+1)-\phi(t_1) \leq 1~.
\label{Eq:ContinuousConstraint}
\end{equation}

\begin{figure}[!ht]
\centering
  \includegraphics[width=10cm]{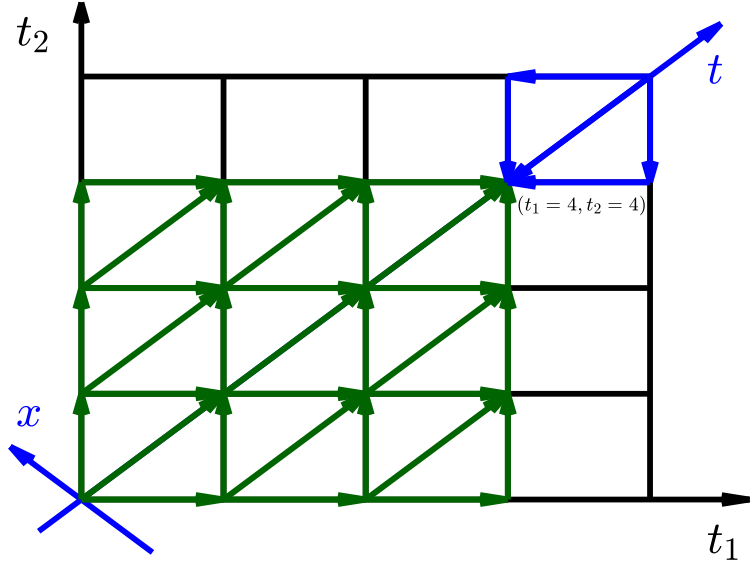}
  \caption{Schematic representation of the TOPS method with time-reversed symmetric weight. The green and blue arrows represent time-forward and time-backward directions, respectively.}
  \label{Fig:TOPS:Schematic:Lattice}
\end{figure}

Based on the transfer matrix introduced by \citet{Sornette-Zhou-2005-QF}, one can transform the coordinates $(t_1,t_2)$ to $(t,x)$ as follows
\begin{equation}
\left\{
  \begin{array}{ccl}
  t = t_2+t_1\\
  x = t_2-t_1,
  \end{array}
\right.
\label{Eq:AxesTransform}
\end{equation}
and obtain the optimal thermal path $\langle x(t) \rangle$ by
\begin{equation}
\langle x(t) \rangle = \sum_{x}x\frac{\overrightarrow{W}(t,x)/\overrightarrow{W}(t) + \overleftarrow{W}(t,x)/\overleftarrow{W}(t)}{2}~,
\label{Eq:ThermalAveragePath TOPS}
\end{equation}
where $\overrightarrow{W}(t,x)$ is the local weight factor for the searching direction from past to future and
\begin{equation}
  \overrightarrow W(t) = \sum_{x} \overrightarrow W(t,x).
\end{equation}
Correspondingly, one can uncover the lead-lag structure between the VIX futures and VIX by quantifying $\langle x(t) \rangle$. If $\langle x(t) \rangle>0$, it represents the VIX futures lead VIX, otherwise the VIX futures lag VIX when $\langle x(t) \rangle<0$. Specifically, the VIX futures and VIX show a contemporaneous movement in price discovery when $\langle x(t) \rangle=0$.
Here, $\overrightarrow W(t,x)/\overrightarrow W(t)$ is the probability for a path to be present on a fixed node $(t,x)$. Following the continuity constraint Eq.~(\ref{Eq:ContinuousConstraint}), a feasible path arriving at $(t_1+1,t_2+1)$ can stem from $(t_1+1,t_2)$ vertically, $(t_1,t_2+1)$ horizontally, or $(t_1,t_2)$ diagonally. Thus, one can calculate the local weights at $(t,x)$ in a recursive way as follow,
\begin{equation}
  \overrightarrow W(t+1,x)=[\overrightarrow W(t,x-1)+\overrightarrow W(t,x+1)+\overrightarrow W(t-1,x)]e^{-\epsilon(t+1,x)/T},
\label{Eq:Recursive:W}
\end{equation}
where $T$ is a parameter controlling the effect of noise. Likewise, $\overleftarrow{W}(t,x)$ represents that the recursive weight process is along the time-backward direction. Fig.~\ref{Fig:TOPS:Schematic:Lattice} illustrates a time-reversed invariant node weight process \citep{Meng-Xu-Zhou-Sornette-2017-QF}, which expounds a feasible path arriving at $(t_1=4,t_2=4)$. In Fig.~\ref{Fig:TOPS:Schematic:Lattice}, green arrows stand for time-forward direction, while blue arrows represent time-reversed direction.

\definecolor{myblackgreen}{rgb}{0,0.390625,0}

\subsection{Self-consistent test of the lead-lag relationships}

It is necessary to conduct self-consistent test for the lead-lag relationships $\langle x(t) \rangle$ determined by the TOPS method. The criteria of the test is whether $X(t-\langle x(t) \rangle)$ and $Y(t)$ exhibit a strong linear dependence. In other words, self-consistent test leads to the following regression
\begin{equation}
  Y(t)=c+a X(t-\langle x(t) \rangle)+\varepsilon(t),
  \label{Eq:Synchronized Self-Consistent}
\end{equation}
where the coefficient $a$ should be significantly different from $0$ for a statistically significant correlation \citep{Meng-Xu-Zhou-Sornette-2017-QF,Xu-Zhou-Sornette-2017-JIFMIM}.

\section{Data description and preliminary analysis}
\label{Sec: Data description}

This study is based on the daily prices of the VIX and VIX futures, which are freely available at the web site of the CBOE. The sample period covers from March 26, 2004 to June 19, 2017 for a total of 3331 observations. Since there exist several VIX futures contracts traded on each day with different maturity date, we need to construct a single continuous price series. Inspired by other research, we splice together the nearest-term contracts (the most actively traded contracts), which are rolled over on the day when the trading volume of the second nearest-term contract exceeds the trading volume of the nearest-term contract \citep{Frijns2016,Zhang2012,Shao2019,Wang-Yang-2010-EE}. Taking into account the effects of using different returns on the lead-lag relationship, we conduct the TOPS method using both the closing and settlement returns for comparative analysis. We refer to the closing and settlement price series of VIX futures as the VXFC and VXFS, respectively.
For the subsequent analysis in this study, logarithmic returns are employed as follows
\begin{equation}
  r(t)=\ln p(t)-\ln p(t-1),
  \label{Eq:LogarithmicReturn}
\end{equation}
where $p(t)$ is the VIX and VIX futures prices  at time $t$.

\begin{figure}[!ht]
	\centering
	\includegraphics[width=16cm]{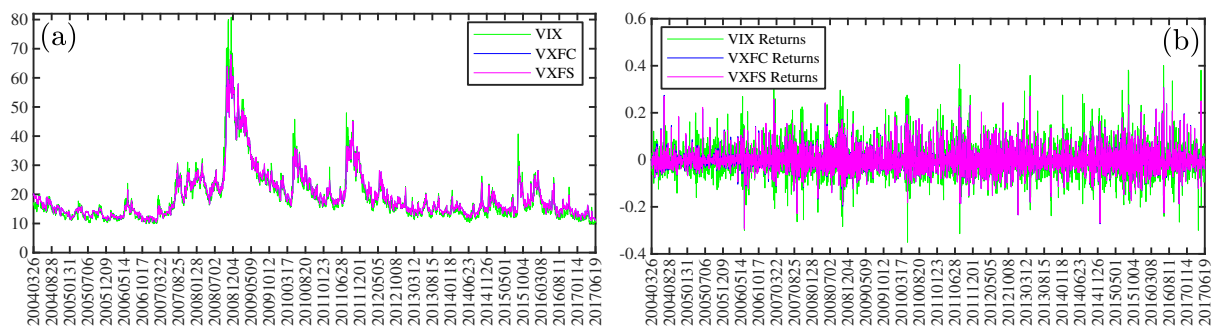}
	\caption{The evolution of the daily prices and returns of the VIX and VIX futures. (a) Prices of the VIX and VIX futures. (b) Returns of the VIX and VIX futures.}
	\label{Fig:Price:Return:Basis}
\end{figure}

\begin{table}[htp]\addtolength{\tabcolsep}{2pt}
\footnotesize
\caption{Summary statistics of the VIX and VIX futures returns.}
\label{Tb:Results:SubSamples}
\medskip
\centering
\begin{threeparttable}
\begin{tabular}{lllllll}
\hline\hline
& VIX & VXFC & VXFS  \\
\hline
\multicolumn{4}{l}{\textit{Panel A : Returns}} \\
 Mean & -0.00015& -0.00019& -0.00019 \\
 Maximum & 0.49601& 0.30696& 0.30625 \\
 Minimum & -0.35059& -0.29094& -0.29480 \\
 Std.Dev & 0.07047& 0.05120& 0.05109 \\
 Skewness & 0.69378& 0.75401& 0.75763 \\
 Kurtosis & 7.15191& 7.16927& 7.22719 \\
 $\rm JB_{\it p-\rm value}$\tnote{a} & 0.00000& 0.00000& 0.00000 \\
 $\rm ADF_{\it p-\rm value}$\tnote{b} & 0.00000& 0.00000& 0.00000 \\
 $\rm ADF_{\it p-\rm value}$\tnote{c} & 0.00000& 0.00000& 0.00000 \\
 \\
\multicolumn{4}{l}{\textit{Panel B: Correlation matrix of returns}} \\
 VIX & 1.00000  \\
 VXFC & 0.78318& 1.00000  \\
 VXFS & 0.78440& 0.99784& 1.00000  \\
\hline\hline
\end{tabular}
\begin{tablenotes}[para,flushleft]
Notes: \\
\item[a]  $\rm JB$ is the Jarque-Berra test of normality, which is distributed as $\chi^2(2)$, and the $\rm JB_{\it p-\rm value}$ is the associated $p$-value. \\
\item[b]  ADF is the Augmented Dickey-Fuller test of unit root, which contains a constant in regression equation, and the $\rm ADF_{\it p-\rm value}$ is the associated $p$-value. \\
\item[c]  ADF is the Augmented Dickey-Fuller test of unit root, which contains a constant and a linear trend in regression equation, and the $\rm ADF_{\it p-\rm value}$ is the associated $p$-value. \\
\end{tablenotes}
\end{threeparttable}
\end{table}

Fig.~\ref{Fig:Price:Return:Basis} plots the daily prices and returns of the VIX and VIX futures. Roughly, we find that the VIX and VIX futures prices move along the same direction for most of the time. Fig.~\ref{Fig:Price:Return:Basis}(a) presents that booming increasing of the VIX and VIX futures in August 2008, May 2010, July 2011 can be attributed to the Global Financial Crisis and European Debt Crisis. And the fundamental reason is the sharp drop of the S\&P 500 index during these periods. Futher, Fig.~\ref{Fig:Price:Return:Basis}(b) shows a apparent volatility clustering in return series. Fig.~\ref{Fig:Price:Return:Basis} indicates that the VXFC is almost in line with the VXFS.

Table~\ref{Tb:Results:SubSamples} reports the summary statistics of the VIX and VIX futures returns. The mean and maximum values of the VIX returns are slightly larger than those of the VIX futures returns, but the minimum value of the VIX futures returns is smaller than that of the VIX returns. In other words, the VIX returns hold more extreme values than the VIX futures returns. As can be seen from standard deviation, the VIX returns keep a larger value. Both returns series present positive skewness and have excess kurtosis. Confirmed by the Jarque-Bera normality tests at the $1\%$ significance level, both return series did not obey normal distribution. Furthermore, when carrying out the ADF unit root tests on the VIX and VIX futures returns, one can reject the presence of a unit root in both returns series at the $1\%$ significance level. Specially, as the different $\rm ADF_{\it p-\rm value}$ reported in Table~\ref{Tb:Results:SubSamples}, we verify  stationarity of both returns series by applying two models \citep{Shao2019,Said-Dickey-1984-Bm}. In addition, there exists a strong correlation between the VIX returns and the VIX futures returns, which is confirmed with a correlation coefficient close to 0.8.

\section{Dynamic lead-lag relationships between the VIX and VIX futures}
\label{Sec: Application}

In this section, the interaction patterns between the VIX and VIX futures returns are investigated by the TOPS method. In order to ensure the comparability of the returns series, we standardized the return series $r(t)$ as follows:
\begin{equation}
R(t)=\frac{r(t)-\bar{r}}{\sigma},
\label{Eq:ReturnNormalization}
\end{equation}
where $\bar{r}$ and $\sigma$ are the mean and standard deviation of return series $r(t)$.

\begin{figure}[!ht]
\centering
  \includegraphics[width=16cm]{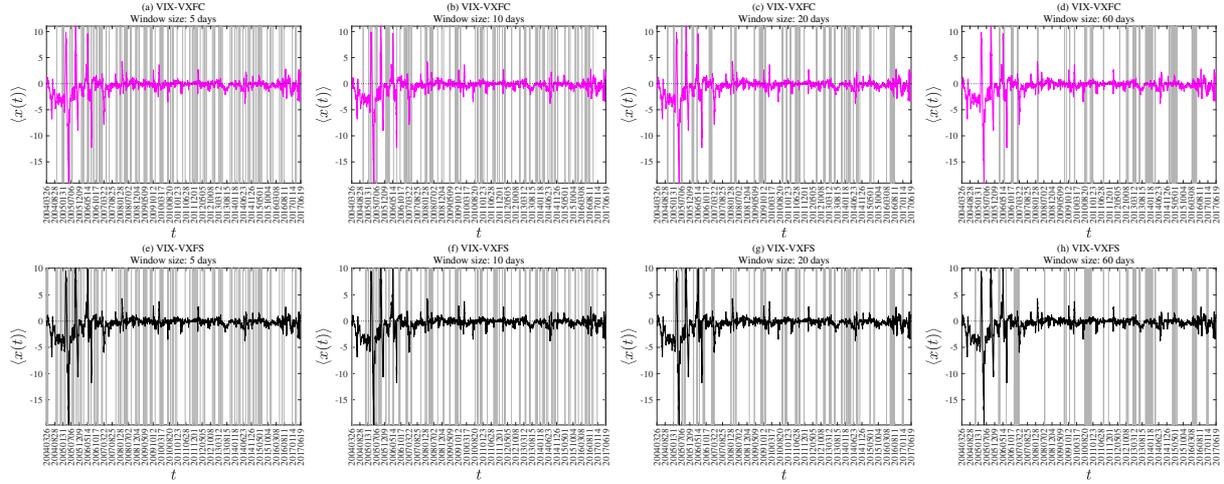}
  \caption{TOPS analysis of the normalized returns $R(t)$ of daily VIX and VIX futures. The first and second rows report the average optimal thermal path $\langle x(t)\rangle$ implemented at $T=2$, which correspond respectively to two pairs of ``VIX vs. VXFC'' and ``VIX vs. VXFS''. The domains in grey represents the times when the self-consistent test is significant at the $5\%$ level.}
  \label{Fig:TOPS:Daily}
\end{figure}

Fig.~\ref{Fig:TOPS:Daily} depicts the average optimal thermal path $\langle x(t)\rangle$ between the normalized returns $R(t)$ of the daily VIX and VIX futures, together with the self-consistent test Eq.~(\ref{Eq:Synchronized Self-Consistent}). Following \citet{Meng-Xu-Zhou-Sornette-2017-QF}, the TOPS path $\langle x(t) \rangle$ is determined through the calculation of lowest free energy among $41 \times 41$ paths of different starting points ($t_1=i_1,t_2=i_2$) and ending points ($t_1=N-i_1,t_2=N-i_2$) for $i_1,i_2=0,1,2,\cdots,30$.
\citet{Meng-Xu-Zhou-Sornette-2017-QF} perform intensive synthetic tests to demonstrate that a temperature close to $T=2$ possesses the best performance for the TOPS analysis. Thus, we
set $T$ to be 2 for capturing the lead-lag structure between the VIX and VIX futures returns. Moreover, there exists no obvious difference among the TOPS path $\langle x(t) \rangle$ with the assignment of $T = 0.5$, 1 and 1.5, respectively. Therefore, Fig.~\ref{Fig:TOPS:Daily} presents a robust average optimal thermal path $\langle x(t)\rangle$. As stated above, a positive $\langle x(t) \rangle$ implies the VIX returns lead the VIX futures returns, and vice versa.

The upper panel of Fig.~\ref{Fig:TOPS:Daily} shows the lead-lag relationships between the VIX and VIX futures based on the closing price. For comparison, the lower panel of Fig.~\ref{Fig:TOPS:Daily} reveals the average optimal thermal path $\langle x(t)\rangle$ between the VIX and VXFS returns. One can observe that there does not arise to be any significant difference in the TOPS path $\langle x(t) \rangle$ whether the VXFC or the VXFS is used. Specially, Fig.~\ref{Fig:TOPS:Daily} unveils a time-varying lead-lag structure between the VIX and VIX futures, which means the lead-lag relationships exist only temporarily.

The empirical evidence provided in Fig.~\ref{Fig:TOPS:Daily} suggests that the VIX leads the VIX futures in the first few years, especially before the introduction of the VIX options on February 24, 2006. Although VIX futures were introduced by the CBOE on March 26, 2004, trades were rather dismal until the introduction of VIX options. In other words, the VIX futures was less informative before February 24, 2006. Thus, the dominance of the VIX in price discovery can be attributed to the less informativeness of the VIX futures. With the launch of VIX ETPs, trading in the VIX futures markets has become dramatically brisk \citep{Fernandez-perez2019}. Meanwhile, one can find that the VIX futures have contributed increasingly to the price discovery process of volatility.
In addition, the TOPS path $\langle x(t)\rangle$ seems to alternate and mainly fluctuates in the range $[-5,5]$ since February 24, 2006.

\begin{figure}[!ht]
\centering
  \includegraphics[width=16cm]{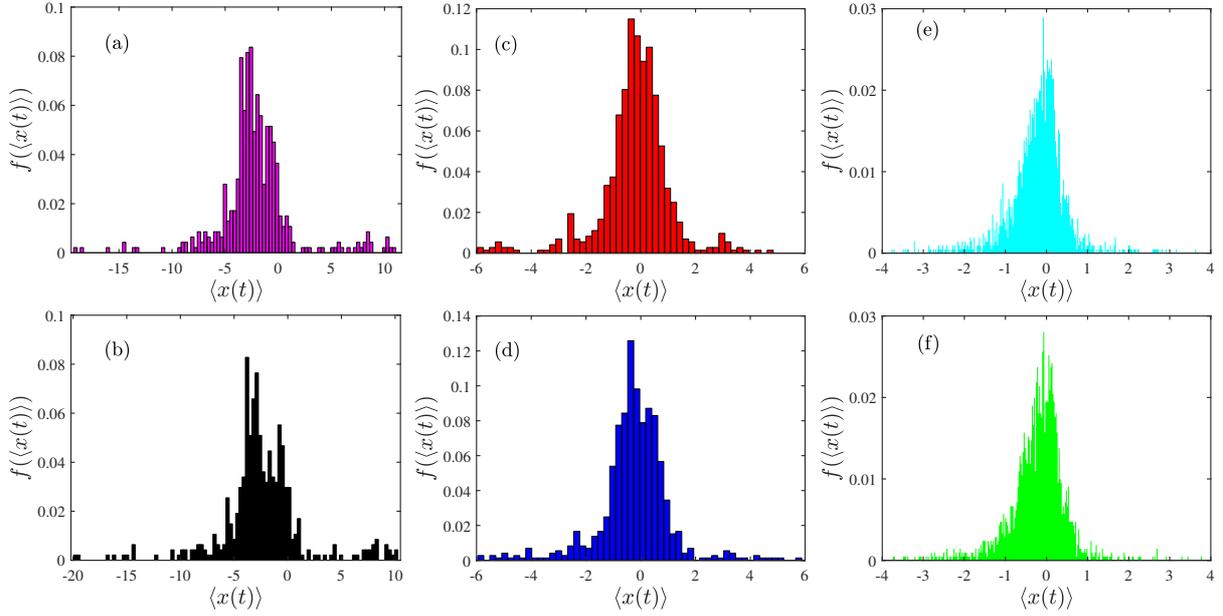}
  \caption{Histogram of TOPS paths $\langle x(t) \rangle$ at $T=2$. (a) Histogram corresponds to the VXFC from March 26, 2004 to February 23, 2006. (b) Histogram corresponds to the VXFS from March 26, 2004 to February 23, 2006. (c) Histogram corresponds to the VXFC from February 24, 2006 to January 28, 2009. (d) Histogram corresponds to the VXFS from February 24, 2006 to January 28, 2009. (e) Histogram corresponds to the VXFC since January 29, 2009. (f) Histogram corresponds to the VXFS since January 29, 2009.}
  \label{Fig:TOPS:Xt:PDF}
\end{figure}

To explicitly assess the lead-lag structure between the VIX and its futures, we conduct further analysis using histogram of $\langle x(t)\rangle$ in Fig.~\ref{Fig:TOPS:Xt:PDF}. For simplicity, we divide the full sample into three distinct phases since the launch of VIX futures on March 26, 2004. Phase 1 began with the launch of VIX futures, which corresponded to period from March 26, 2004 to February 23, 2006. Phase 2 covers the period from  February 24, 2006 to January 28, 2009, while third sub-sample keeps the rest period. All three phases have witnessed an uptake in the VIX futures trading activity. Corresponding to the VXFC, Fig.~\ref{Fig:TOPS:Xt:PDF}(a)(c)(e) illustrate the probability distribution of $\langle x(t)\rangle$ for Phase 1, Phase 2 and Phase 3, respectively. Fig.~\ref{Fig:TOPS:Xt:PDF}(a)(c)(e) show that the percentages of negative values of $\langle x(t)\rangle$ are $87.77\%$, $55.82\%$, $62.04\%$ for the VXFC returns series, respectively. Fig.~\ref{Fig:TOPS:Xt:PDF}(a) presents that $\langle x(t)\rangle$ mainly concentrates in the range $[-5, 0]$ with the mean value $\overline{\langle x(t)\rangle}=-2.15$. The TOPS path $\langle x(t)\rangle$ of Phase 1 indicates that the VIX returns lead its futures returns by 0-5 days in most of the time. Fig.~\ref{Fig:TOPS:Xt:PDF}(c) shows that $\langle x(t)\rangle$ mainly fluctuates in the interval $[-2, 2]$ with the mean value $\overline{\langle x(t)\rangle}=-0.23$. The histogram of $\langle x(t)\rangle$ in Phase 2 demonstrates that the VIX and its futures affect each other to about the same extent. Fig.~\ref{Fig:TOPS:Xt:PDF}(e) reveals that the $\langle x(t)\rangle$ mainly varies in the range $[-1, 1]$ with the mean value $\overline{\langle x(t)\rangle}=-0.21$. Additionally, the evolution of $\langle x(t)\rangle$ from Phase 1 to Phase 3 implies that the VIX futures have been increasingly more important in the price discovery. When it comes to the histogram of $\langle x(t)\rangle$ for the VXFS, one can observe extremely similar characteristics compared with the VXFC. Fig.~\ref{Fig:TOPS:Xt:PDF}(b)(d)(f) show that the percentages of negative values of $\langle x(t)\rangle$ are $86.62\%$, $60.03\%$ and $61.62\%$, while the mean value $\overline{\langle x(t)\rangle}$ are -2.15, -0.23 and -0.21, respectively. Overall, neither the VIX nor its futures
plays a predominant role in the price discovery process since the launch of aforementioned VIX options and VIX ETPs.

We also implement the self-consistent test defined by equation Eq.~(\ref{Eq:Synchronized Self-Consistent}), within moving windows with sizes of 5 to 60 days. The grey shades in Fig.~\ref{Fig:TOPS:Daily} represent the times when the consistency test is significant at the $5\%$ level. For both the VXFC and VXFS returns, Fig.~\ref{Fig:TOPS:Daily} unveils that regression test of Eq.~(\ref{Eq:Synchronized Self-Consistent}) is significant in most of the time during the whole period.

\section{Conclusion}
\label{Sec: Conclustion}

The purpose of this work is to explicitly examine interaction patterns between the VIX and VIX futures for the period March 26, 2004 to June 19, 2017. Based on the TOPS method, we carry out analysis of the lead-lag structure between the VIX and VIX futures. Further, we confirm the robustness of the lead-lag structure by applying the self-consist test.

The empirical results suggest several primary findings. First, there exists a time-dependent lead-lag relationships between the VIX and VIX futures in most of the time. Second, the VIX lead its futures in the first few years. Specially, the VIX futures lagged the VIX by less than 5 days in most time of the Phase 1. We conjecture that this might be because of the lower volume of trading in the VIX futures markets. We further find that the VIX futures become much more important over time, especially after the introduction of the VIX options in 2006 and the VIX ETPs in 2009. In other words, the lead-lag dynamics seem to alternate since February 24, 2006, with sometimes the VIX leading its futures, while at other times the VIX futures leads the VIX. This could be explained by the launch of VIX derivatives which has caused the increment in the informativeness of the VIX futures markets.
Our results are of great importance to the research of causality between the VIX and its futures \citep{Frijns2016,Zhang2012,Bollen2017,Chen2017}.

\section*{Acknowledgements}

This work was supported by Peak Discipline Construction Project of Education at East China Normal University, the National Natural Science Foundation of China (Grant No. 11805119).

\end{document}